\documentclass[a4paper]{PoS}

\newcommand{\Ts}{T_{\rm S}}
\newcommand{\Tk}{T_{\rm K}}

\newcommand{\nf}{x_{\rm HI}}
\newcommand{\avenf}{\bar{x}_{\rm HI}}

\newcommand{\lya}{Ly$\alpha$}
\newcommand{\lyb}{Ly$\beta$}

\newcommand{\Msun}{M_\odot}
\newcommand{\Tvir}{T_{\rm vir}}

\newcommand{\Tcmb}{T_\gamma}
\newcommand{\delT}{\delta T_b}

\newcommand{\delNL}{\delta_{\rm nl}}

\newcommand\lsim{\mathrel{\rlap{\lower4pt\hbox{\hskip1pt$\sim$}}
        \raise1pt\hbox{$<$}}}
\newcommand\gsim{\mathrel{\rlap{\lower4pt\hbox{\hskip1pt$\sim$}}
        \raise1pt\hbox{$>$}}}
\def\myputfigure#1#2#3#4#5%
{\vskip#5pt\makebox[0pt]{\hskip#2in
\includegraphics[width=#3\textwidth]{#1}}\vskip#4pt\hfill}

\title{How the first generations of luminous baryons established the X-ray and UV backgrounds}

\ShortTitle{How the first generations of luminous baryons established the X-ray and UV backgrounds}

\author{\speaker{Andrei Mesinger}%
        \thanks{Hubble Fellow}\\
       Princeton University\\
       E-mail: \email{mesinger@astro.princeton.edu}}

\abstract{The first generations of astrophysical objects made a substantial impact on our Universe with their radiation.  X-rays from the first sources, with large mean free paths, likely quickly heated the intergalactic medium (IGM).  The second generation of 21cm instruments can provide a unique view into this early epoch.  The early stages of reionization likely followed, driven by so-called "minihalos", i.e. molecularly-cooled halos.  These small halos were susceptible to complex feedback mechanisms, especially from the soft-UV background which preceded reionization, resulting in complex and possibly extended early stages of reionization.  When atomically-cooled galaxies emerged as the dominant ionizers, reionization could proceed more rapidly, with these being less sensitive to radiative feedback than previously thought. Reionization could have slowed in the final stages when the ionized bubbles grew larger than the separation of Lyman limit absorption systems (LLSs).  The final stages likely involved the photo-evaporation of LLSs, which by then regulated the rise of the UV background.  I discuss the theoretical underpinnings of this narrative, as well as how future 21cm observations may help shed light on the outstanding uncertainties.}

\FullConference{Cosmic Radiation Fields: Sources in the early Universe\\
		 November 9-12, 2010\\
		 Desy, Germany}

\begin{document}

\section{Intro}
The dawn of the first astrophysical objects and the eventual reionization of the Universe is of fundamental importance, offering insight into many astrophysical processes.  Also, since observational surveys only pick out the rarest, most luminous objects, cosmic radiation fields and their effects on the intergalactic medium (IGM) may our only way of studying the majority of our galactic ancestors.  Recent theoretical and observational advances are starting to shed light on a complex and extended reionization epoch.  Here I sketch-out three likely stages of this process.  I briefly discuss the potential of the redshifted 21cm signal to probe not only reionization, but also the pre-reionization epochs.  However, interpreting the upcoming observations requires efficient modeling tools, such as the recently developed 21cmFAST.

\section{{\it \bf Once upon a time...} the story of cosmic dawn and reionization}

\subsection{Early stages}

Hierarchal structure formation implies that the first astrophysical objects are likely hosted by so-called minihalos (with virial temperatures $\Tvir \lsim 10^4$ K).  Cooling via atomic hydrogen is inefficient at such low temperatures, so baryons condense into these halos via the molecular hydrogen (H$_2$) cooling channel.  Without metals to aid in cooling and fragmentation, the first generations of stars (so-called PopIII stars) are likely massive (with masses of $\sim100\Msun$; \cite{ABN02, BCL02, YOH08}), yet short-lived (with lifetimes of a few Myr; \cite{Schaerer02, Schaerer03}). 

PopIII stars are likely to have very different properties from ``normal'' PopII stars, with harder spectra, a factor of ten times more ionizing photons per baryon \cite{Schaerer02}, and various exotic, mass-dependent fates (e.g., \cite{Heger03}). 
  Accurately determining their properties, formation efficiency and initial mass functions (IMFs) requires detailed numerical simulations and costly explorations of parameter space (e.g., \cite{TAO09}).

These prehistoric giants have complex interactions with their surroundings, since star formation in such small-mass halos is susceptible to feedback mechanisms.  Modeling these requires introducing parameterized prescriptions for stellar properties and extending simulations to larger, $\sim1$ Mpc scales. Feedback processes can come in three forms: (1) {\it mechanical:} supernovae (SNe) can blow out gas from the surrounding shallow potential well, delaying local star formation until gas is re-accreted (e.g., \cite{Whalen08, WA07}); (2) {\it chemical:} evolution in the IMF depends on metal enrichment \cite{TM08, Smith09, SO10}, which is very inhomogeneous making it necessary  to model the enrichment on large scales (e.g., \cite{TFS07}); and (3) {\it radiative}: radiation in various bands can either promote (positive feedback) or suppress (negative feedback) future star formation.

Because it acts on very large scales and involves a large dynamic range in energy, it is likely that radiative feedback is the most pivotal and complex of the feedback mechanisms.  Positive radiative feedback can result  when the enhanced free-electron fraction from ionizing photons (e.g., \cite{OH02}) or hydrodynamical shocks \cite{SK87} catalyzes the formation of H$_2$, thereby enhancing the H$_2$ cooling channel.  Negative feedback can result from heating by ionizing radiation which can photo-evaporate gas in low-mass halos (e.g., \cite{Efstathiou92}).  Also, an active background of Lyman-Werner (LW) radiation (with photon energies in the 11.18--13.6eV range) can dissociate H$_2$, thus decreasing the gas's cooling capabilities (e.g., \cite{HRL97}).

The first astrophysical objects can emit radiation in various bands.  X-rays, generated from accretion onto remnant black holes or massive X-ray binaries, have long mean free paths, but their feedback effects seem to be mild \cite{MBA03, KM05}.  However, they can be responsible for early IGM heating and some degree of reionization (e.g., \cite{DHL04, RMZ08, MFC10}).  Feedback resulting from a LW background is by definition negative, and can delay star formation in low mass halos, depending on the strength of the background \cite{MBA01, MBA03, Yoshida03, ON08}.
Also depending on its strength, a transient (due to the short lifetimes of PopIII stars) ionizing UV background (UVB) can have various feedback regimes \cite{OH03, MBH06, MBH09, WA07, Whalen08}, especially in conjunction with the negative feedback effects of a LW background \cite{MBH06, MBH09, WA07}.  Despite this complexity, it is likely that the final suppression of star formation inside minihalos was regulated by the build-up of persistent UV or LW backgrounds before the bulk of reionization \cite{HAR00, HB06}.  However, the complexity of radiative feedback and/or an early X-ray background could precipitate an extended early reionization epoch, which could allow a significant amount of reionization to occur at late times ($z\sim$6), and still be consistent with constraints from the Wilkinson Microwave Anisotropy Probe (WMAP) \cite{Larson10}.

\subsection{Middle stages}

\begin{figure*}[t]
%\vspace{-2cm}
{
\vspace{-0\baselineskip}
\begin{center}
\includegraphics[width=16.cm]{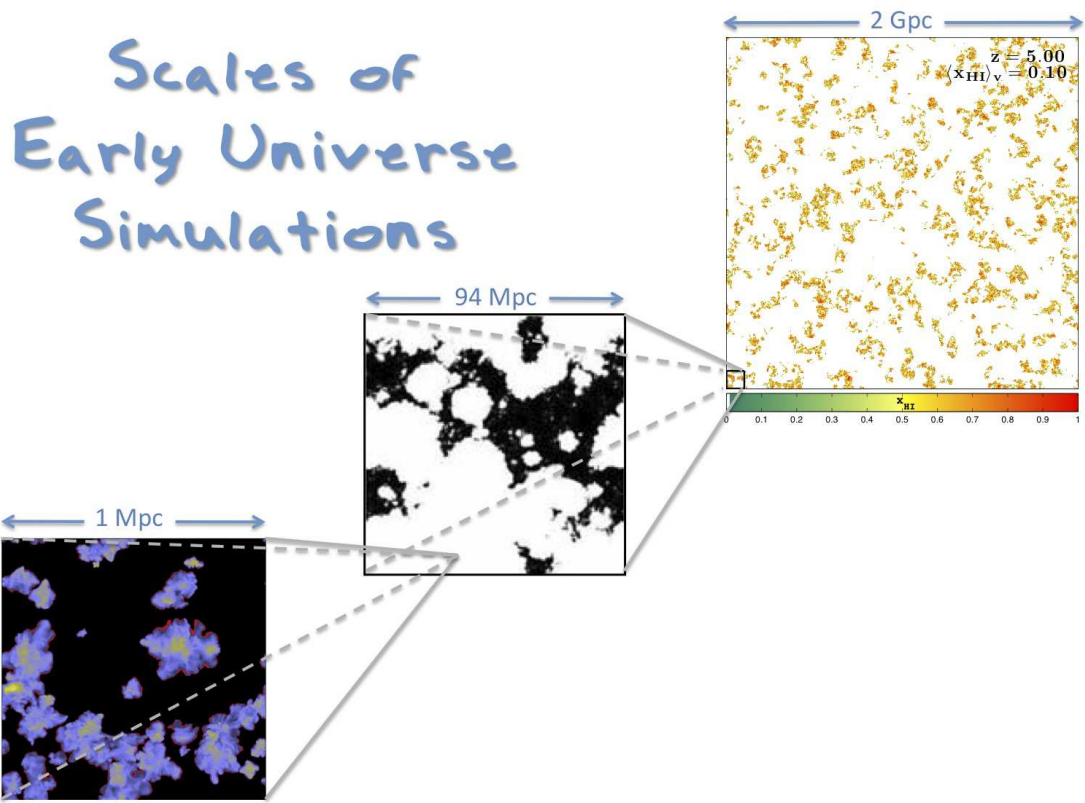}
\end{center}
}
\vspace{-2.5\baselineskip}
\caption{
{\footnotesize Scales of some early Universe simulations, with box sizes indicated in comoving units.  Slices correspond to metallicity maps from \cite{Wise10}, and ionization structure from \cite{McQuinn07} and \cite{Mesinger10} ({\it left to right}). The left and middle panels roughly correspond to the current maximum box sizes of state-of-the-art numerical simulations of the first stars and the middle stages of reionization, respectively.  This figure highlights the daunting dynamic range of cosmological processes such as reionization. In principle, one should continue ``zooming-in'' down to stellar scales (e.g., \cite{Schaerer02, Heger03}).
\label{fig:scales}}}
%\vspace{-1\baselineskip}
\end{figure*}

After the initial, ``fireworks'' stage of PopIII-driven reionization, longer-lived stars begin to establish persistent UVBs.  Reionization proceeds in an ``inside-out'' fashion on large-scales, with biased sources driving HII regions to tens of Mpc in size (e.g., \cite{FZH04, FO05, Zahn10}).  Larger, atomically-cooled halos ($\Tvir\gsim10^4$ K) start dominating the ionization budget.  These are likely too faint to image directly with current or upcoming instruments (e.g., \cite{SFD10, WL10}), although we can get close with gravitational lensing surveys (e.g., \cite{Stark07, MF08LAE, WYW11}).
  Likewise, because of the inside-out nature of reionization, most astrophysical objects during the epoch of reionization find themselves 
inside large HII regions, largely unaffected by the neutral IGM even when it still occupies a substantial volume filling fraction. Thus most of the current observational probes of reionization (derived from quasar proximity regions, damping wings in quasar and gamma ray burst (GRB) spectra, \lya\ emitter (LAE) number density and clustering properties; see below) are most effective back when the Universe is predominantly neutral.

Aside from these observational challenges, there are substantial theoretical hurdles to overcome when studying the middle stages of reionization.  Ionization structures during this epoch can span volumes over four magnitudes larger than the first-stars simulations discussed in the previous section (see Fig. 1).  To statistically capture the middle stages of reionization, simulation boxes need to be hundreds of Mpc in length, yet still be able to account for ionizing radiation from small-mass halos.  The required dynamic range is daunting, and state-of-the-art simulations are forced to ignore minihalos and photon sinks, while also using approximate prescriptions to assign ionizing luminosities to source halos (see the recent review in \cite{TG09}).  More approximate techniques have recently been developed in order to overcome the challenge of such a large range of scales, 
including sub-grid \cite{KGH07,McQuinn07} and semi-numerical \cite{Zahn07, MF07, GW08, Alvarez09, CHR09, Thomas09, MFC10} models.

Additionally, our poor understanding of the early Universe means that large explorations of parameter space are required for any robust conclusion.  Analytic models have been very useful in this respect, predicting that HII morphology is not very sensitive to redshift, when normalized to the same mean neutral fraction, $\avenf$, and also that sources are more important that ionizing sinks in this regime \cite{FZH04, FO05}.  These predictions were confirmed by the first numerical parameter studies in \cite{McQuinn07}.  Another parameter study, \cite{MD08}, combined 1D hydrodynamical collapse simulations with large-scale 3D semi-numerical simulations of the UVB during reionization, and concluded that radiative feedback does not play a major role during this epoch.  Results of such parameter studies allow us to simplify various theoretical problems.

\subsection{Final stages}

As reionization progresses, HII regions grow to be larger than the mean free path of ionizing photons through them, requiring even larger simulation boxes (see Fig. 1).  In this final regime, sinks of ionizing photons, such as Lyman limit systems (LLSs) could regulate the progress of reionization \cite{FM09}, and its morphological structure \cite{FO05, CHR09}.  These absorption systems are likely found in the filaments of the recently-ionized IGM, where the UVB is the weakest \cite{Crociani10}.  The final ``overlap'' stages of reionization could be delayed until either absorbers get photo-evaporated (e.g. \cite{ISR05}), or the remaining neutral islands get ionized from the inside.  The time-scales for these processes are of order $\Delta z \sim 1$.  This, combined with the large cosmic variance during this epoch (e.g., \cite{Mesinger10}), suggests that the overlap stage is much more gradual than was initially predicted by small-box simulations (e.g., \cite{Gnedin00a}).  It is also unlikely that these final stages leave a strong observational imprint, such as a sharp rise in the UVB \cite{FM09}.  Despite popular belief, current observations cannot rule-out a very late overlap, even at $z\sim5$ \cite{Lidz07, Mesinger10, MMF11}.

\subsection{Observational constraints}

Several model-dependent constraints on $\avenf$ at $z\sim6$ have been 
derived from various astrophysical probes such as: (1) the size of the proximity 
zone around quasars (\cite{WL04_nf,Carilli10}, but see \cite{MHC04,BH07_quasars,Maselli07});
 (2) a claimed detection of damping wing absorption from neutral IGM in quasar 
spectra 
(\cite{MH04,MH07}, but see \cite{MF08damp});
 (3) the {\it non}-detection of intergalactic damping wing absorption in a 
GRB spectrum 
(\cite{Totani06}, but see \cite{McQuinn08});
 (4) the number density and clustering of Ly$\alpha$ emitters 
(\cite{MR04,HC05,FZH06,Kashikawa06,McQuinn07LAE}, but see
\cite{DWH07,MF08LAE,Iliev08}); (5) the length of dark gaps in quasar spectra (\cite{Gallerani08}, but see \S 3.3 in \cite{Mesinger10}).
Additionally, a direct, nearly model-independent upper limit on $\avenf$ was recently derived from the dark covering fraction in the \lya\ and \lyb\ forests \cite{MMF11}.

However, constraining the ionization state of the high-redshift Universe is very difficult with current observations, since they are not direct probes of HII morphology.  For example, the Lyman forests begin to saturate at $z\gsim5$, becoming difficult to interpret (e.g., \cite{Songaila04, Fan06, BRS07, Lidz07, Mesinger10}).  Similarly, the dominant population of ionizing galaxies is $\sim$ two orders of magnitude fainter than current detection limits; and in any case, galaxies offer a very biased tracer of the cosmic radiation fields.  Luckily, there is a plethora of upcoming observations which should help advance our understanding:
 21cm tomography, high-redshift IR spectra, wide-field LAE surveys, and improved E-mode CMB polarization power spectra.
  Of these, arguably the 21cm line from neutral hydrogen provides the greatest potential for directly studying reionization in the near term.

\section{The 21cm frontier and 21cm{\it \bf FAST}}
 
The cosmological 21cm signal uses the CMB as a back-light.  The offset of the 21cm brightness temperature from the CMB temperature, $\Tcmb$, along a line of sight (LOS) at observed frequency $\nu$, can be written as (c.f. \cite{FOB06}):
\begin{equation}
\label{eq:delT}
\delT(\nu) = 27 \nf \left(1 - \frac{\Tcmb}{\Ts} \right) (1+\delNL) \left(\frac{H}{dv_r/dr + H}\right) \left( \frac{1+z}{10} \frac{0.15}{\Omega_{\rm M} h^2}\right)^{1/2} \left( \frac{\Omega_b h^2}{0.023} \right) {\rm mK},
\end{equation}
\noindent where $\nf$ is the hydrogen neutral fraction, $T_S$ is the gas spin temperature, $\tau_{\nu_0}$ is the optical depth at the 21cm frequency $\nu_0$, $\delNL({\bf x}, z) \equiv \rho/\bar{\rho} - 1$ is the evolved (Eulerian) density contrast, $H(z)$ is the Hubble parameter, $dv_r/dr$ is the comoving gradient of the line of sight component of the comoving velocity, and all quantities are evaluated at redshift $z=\nu_0/\nu - 1$.  The above equation also assumes that $dv_r/dr \ll H$, which is generally true for the pertinent redshifts and scales.

The redshifted 21cm line carries with it much information, containing both astrophysical ($\nf$ and $\Ts$) and cosmological (last four terms in eq. \ref{eq:delT}) terms (see \cite{Furlanetto09_21cmastro, Furlanetto09_21cmcosmo}).  However, in order to make sense of the upcoming observations, one must disentangle the various components.  Considering the scales and uncertainties involved, this is a daunting task.

Recently ``semi-numerical'' tools have proven invaluable in overcoming these obstacles.  These simulations use more approximate physics than numerical simulations, but are much faster.  In particular, the recent launch of 21cmFAST \cite{MFC10} (http://www.astro.princeton.edu/$\sim$mesinger/Sim) provided the astronomical community a publicly-available tool specialized for fast, large-scale simulations of the 3D cosmological 21cm signal.  For example, a 256$^3$ realization of reionization with 21cmFAST is obtained in just a few minutes on a single CPU, with every step agreeing with cosmological hydrodynamic simulations into the quasi-non-linear regime \cite{MF07, Zahn10, MFC10}.  Such speeds facilitate rapid explorations of the vast parameter space, necessary for interpreting upcoming observations.

Most studies thus far have focused on the reionization term, $\nf$, making various simplifying assumptions in eq. \ref{eq:delT}.  The 21cm signal during reionization can indeed tell us much about the relevant physical processes and intergalactic radiation fields.  However, the 21cm signal also offers us a precious glimpse into even earlier epochs, via the spin temperature, $\Ts$, term.

The spin temperature can be written as (e.g. \cite{FOB06}):
\begin{equation}
\label{eq:Ts}
\Ts^{-1} = \frac{ \Tcmb^{-1} + x_\alpha T_\alpha^{-1} + x_c \Tk^{-1} }{1 + x_\alpha + x_c}
\end{equation}
\noindent where $\Tk$ is the kinetic temperature of the gas, and $T_\alpha$ is the color temperature, which is closely coupled to the kinetic gas temperature, $T_\alpha \approx \Tk$ \cite{Field59}.  From this equation, one can see that the spin temperature interpolates between the CMB and the gas temperature.  It does this through two coupling coefficients: (1) the collisional coupling coefficient, $x_c$, which requires high densities and is effective in the IGM at $z\gsim40$; and (2) the Wouthuysen-Field (\cite{Wouthuysen52, Field58}; WF) coupling coefficient which uses the \lya\ radiation background and is effective soon after the first sources ignite.  If either coefficients is high, the spin temperature approaches the kinetic temperature of the gas.  Otherwise, the spin temperature approaches the CMB temperature and there is no signal.

The CMB temperature decreases as $\propto (1+z)$.  After decoupling from the CMB, the gas kinetic temperature decreases as $\propto (1+z)^2$, before being heated, most likely with X-rays from the first sources (see \cite{Furlanetto06} and the discussion therein).  Likely before the bulk of reionization, the gas is heated to temperatures far beyond the CMB, $\Tk \gg \Tcmb$, and the spin temperature term in eq. \ref{eq:delT} no longer contributes to the signal.  When and where $\Ts < \Tcmb$, the 21cm signal is seen in absorption, and when/where $\Ts > \Tcmb$, the signal is seen in emission.

Therefore, the 21cm signal and its inhomogeneity could be a powerful probe, not only of reionization, but also of the \lya\ background, IGM heating, and cosmology during the collisionally coupled regime before the first sources ignite.  Another probe of cosmology would be available at lower redshifts (making it easier to detect) if heating completes before reionization.  A movie of 21cm evolution from $z=250$, computed with 21cmFAST, is available at http://www.astro.princeton.edu/
$\sim$mesinger/21cm\_Movie.html, together with an explanation of the various interesting epochs.

\section{Conclusions}

\begin{itemize}
\item Milestones such as reionization are likely the only practical way of studying the primordial zoo of astrophysical objects in the near future.\\
\vspace{-0.8cm}
\item Reionization is likely extended, going through various stages.\\
\vspace{-0.8cm}
\item The cosmological 21cm signal is very rich in information, containing both cosmological and astrophysical components.\\
\vspace{-0.8cm}
\item The range of scales and unknown parameter space is enormous.\\
\vspace{-0.8cm}
\item We need efficient modeling tools, such as 21cmFAST, to make sense of the upcoming observations.\\
\vspace{-0.8cm}
\item We are living in exciting times!
\end{itemize}

\noindent {\large \bf Acknowledgments}\\
I would like to thank all of my collaborators who contributed to the science discussed here.  Support for this work was provided by NASA through Hubble Fellowship grant HST-HF-51245.01-A, awarded by the Space Telescope Science Institute, which is operated by the Association of Universities for Research in Astronomy, Inc., for NASA, under contract NAS 5-26555.

\bibliographystyle{abbrv}
\bibliography{ms}

\end{document}